%% file: paper.tex
\documentclass[runningheads]{llncs}
\usepackage{graphicx}
\usepackage{pgfplots}
\usepackage{layouts}

\usepackage{hyperref}
\usepackage{url}
\usepackage[strings]{underscore}
\usepackage{xcolor}
\usepackage{multirow}
\usepackage{authblk}

\begin{document}
\title{Improving BERT-based Query-by-Document Retrieval with Multi-Task Optimization}

\titlerunning{Multi-task optimization for BERT-based QBD retrieval}
%

\author{Amin Abolghasemi\inst{1} 
\and
Suzan Verberne\inst{1} 
\and 
Leif Azzopardi\inst{2}}

\authorrunning{A. Abolghasemi, S. Verberne, L. Azzopardi}
\institute{Leiden University, The Netherlands\\
\email{\{m.a.abolghasemi,s.verberne\}@liacs.leidenuniv.nl}
\and
University of Strathclyde, UK\\
\email{leif.azzopardi@strath.ac.uk}}

\maketitle              
\setcounter{footnote}{0}%
\begin{abstract}
Query-by-document (QBD) retrieval is an Information Retrieval task in which a seed document acts as the query and the goal is to retrieve related documents -- it is particular common in professional search tasks.  
In this work we improve the retrieval effectiveness of the BERT re-ranker, proposing an extension to its fine-tuning step to better exploit the context of queries. 
To this end, we use an additional document-level representation learning objective besides the ranking objective when fine-tuning the BERT re-ranker. 
Our experiments on two QBD retrieval benchmarks show that the proposed multi-task optimization significantly improves the ranking effectiveness without changing the BERT re-ranker or using additional training samples. In future work, the generalizability of our approach to other retrieval tasks should be further investigated.

\end{abstract}

\input{introduction.tex}

\input{background}
\input{methodology}
\input{results}

\section{Conclusion}
This paper shows that it is possible to improve the BERT cross-encoder re-ranker quality using multi-task optimization with an auxiliary representation learning task.
We showed that the resulting model named MTFT-BERT re-ranker obtains consistently better retrieval quality than the original BERT re-ranker using the same training instances and structure.
While our focus was on query-by-document retrieval in professional search domains (legal and academic), as a future work, it would be interesting to study the effectiveness of MTFT-BERT re-ranker in other retrieval tasks where we have shorter queries.

\section{Acknowledgments}
This work is funded by the DoSSIER project under European Union’s Horizon 2020 research and innovation program, Marie Skłodowska-Curie grant agreement No. 860721.

%

\bibliographystyle{splncs04}
\bibliography{references.bib}
\end{document}

%% file: introduction.tex
\section{Introduction}
Query by document (QBD) \cite{yang2009QBD,eugeneyang2018QBD}, is a widely-used practice across professional, domain-specific retrieval tasks \cite{russell2018information,verberne2019first} such as scientific literature retrieval~\cite{mysore2021csfcube,cohan_specter_2020}, legal case law retrieval \cite{althammer2022parm,Askari2021CombiningLA,coliee2020rabelo,Shao2020bertpli}, and patent prior art retrieval~\cite{Fujii2007patentretrieval,piroi2019multilingual}. 
In these tasks, the user's information need is based on a seed document of the same type as the documents in the collection. 
Taking a document as query results in long queries, which can potentially express more complex information needs and provide more context for ranking models~\cite{huston2010evaluatingverbose}. 
Transformer-based ranking models have proven to be highly effective at taking advantage of context~\cite{dai2019deeper,dai2020context,macavaney2019cedr}, but the long query documents pose challenges because of the maximum input length for BERT-based ranking models. 
Recent work showed that transformer-based models which handle longer input sequences are not necessarily more effective when being used in retrieval tasks on long texts~\cite{Askari2021CombiningLA}. 
We, therefore, direct our research towards improving retrieval effectiveness while acting within the input length limitation of ranking models based on large scale pre-trained BERT models~\cite{devlin-etal-2019-bert}.
We posit that the representations learned during pre-training have been tailored toward smaller sequences of text -- and additional tuning the language models to better represent the documents in this specific domain and query setting, could lead to improvements in ranking. 

To investigate this, we first focus on the task of Case Law Retrieval (i.e. given a legal case find the related cases), and employ multi-task optimisation to improve the standard BERT-based cross-encoder ranking model \cite{Humeau2020Poly-encoders:} for QBD retrieval. We then explore the generalizability of our approach by evaluating our approach on four QBD retrieval tasks in the academic domain. 
Our approach draws upon multi-task learning to rank -- where a shared structure across auxiliary, related tasks is used~\cite{ahmad2018multi,liu-etal-2019-multi,qu2020openrankread,cheng2021jointpersonalized}. Specifically, in our method, we employ document-level representation learning as an auxiliary objective for multi-task fine-tuning (MTFT) of the BERT re-ranker. 
To our knowledge, there is no prior work on using representation learning directly as an auxiliary task for fine-tuning a BERT re-ranker.
We show that optimizing the re-ranker jointly with document-level representation learning leads to consistently higher ranking effectiveness over the state-of-the-art with greater efficiency i.e., with the same training instances on the same architecture.

%% file: background.tex
\section{Preliminaries}
\paragraph{BERT-based Ranking.}
\label{bert-based-ranking-section}
Pre-trained transformer-based language models~\cite{devlin-etal-2019-bert} have shown significant improvement in ranking tasks \cite{dai2019deeper,dai2020context,macavaney2019cedr,lin2021pretrained}. 
In this work, we use the BERT re-ranker proposed by Nogueira and Cho~\cite{nogueira2019passage}, which is 
a pre-trained BERT model followed by a projection layer $W_{p}$ on top of its $[CLS]$ token final hidden states. 
The BERT re-ranker, which is a cross-encoder neural ranking model, uses the concatenation of a query and candidate document as the input to a fine-tuned pre-trained BERT model. The output of the model is used to indicate the relevance score $s$ of the document $d$ for the input query $q$, such that:
\begin{equation}
\label{ranking-score-equ}
    s(q,\; d) = BERT\;([CLS]\;  q\;  [SEP]\;  d\;  [SEP])_{[CLS]} * W_{p}
\end{equation}

\paragraph{BERT-based Representation Learning.}
\label{bert-based-rep-learn}
BERT was originally pre-trained on two tasks, namely Masked Language Modeling and Next Sentence Prediction \cite{devlin-etal-2019-bert}. These tasks, however, are not meant to optimize the network for document-level information representation \cite{cohan_specter_2020} which may make the model less effective in representation-focused \cite{guo2016drmm} downstream tasks \cite{dai2019deeper}. Previous works have shown that leveraging a Siamese or triplet network structure for fine-tuning BERT could optimize the model for 
document-level representation \cite{cohan_specter_2020}. 
Following Devlin et al \cite{devlin-etal-2019-bert}, we use the final hidden state corresponding to the $[CLS]$ token to encode the query $q$ and the document $d$ into their representations $r_{q}$, and $r_{d}$:
\begin{equation}
\label{representation-equ}
    r_q = BERT(\;[CLS]\; q\; [SEP]\;)_{[CLS]}\;\;\;\; r_d = BERT(\;[CLS]\; d\; [SEP]\;)_{[CLS]}
\end{equation}

\paragraph{Pairwise Ranking Loss.}
In Learning-To-Rank tasks, pairwise loss minimizes the average number of pairwise errors in a ranked list~\cite{burges_learning_2005,ltrfrompairwise2007}. 
Here, we aim to optimize the BERT re-ranker with a pairwise cross-entropy softmax loss function~\cite{burges_learning_2005}:
\begin{equation}
\label{rank-loss}
l_{rank} =  - log \frac{e^{score(q, d^+)}} {e^{score(q, d^+)}+e^{score(q, d^-)}}
\end{equation}
\noindent where the $score$ function represents the degree of relevance between a query and a document computed as described in Eq.~\ref{ranking-score-equ}. 
In fact, this pairwise loss frames the ranking task as a binary classification problem in which, given a query ($q$) and a pair of relevant ($d^+$) and non-relevant ($d^-$) documents, the fine-tuned ranking model predicts the relevant one. 
However, at inference time the model is used as a point-wise score function.

\paragraph{Triplet Representation Learning Loss.}
In the context of representation learning with pre-trained transformers, a triplet loss function fine-tunes the weights of the model such that given an anchor query $q$, the representations of the query $r_q$ and the document $r_d$ (obtained as described in Eq.~\ref{representation-equ}) are closer for a relevant document $d^{+}$ than for a non-relevant document $d^{-}$:
\begin{equation}
\label{rep-loss}
l_{representation} =  max\{ (\;f(r_q,\; r_{d^{+}})\;-\;f(r_q,\; r_{d^{-}})\;+\;margin\;),\; 0\}
\end{equation}

Here, $f$ indicates a distance metric and and $margin$ ensures that $d^{+}$ is at least $margin$ closer to $q$ than $d^{-}$ \cite{reimers-gurevych-2019-sentence}.

%% file: methodology.tex

\section{Multi-task fine-tuning of the BERT re-ranker}

        \begin{figure}[t]
            \centering
            \includegraphics[scale=0.12]{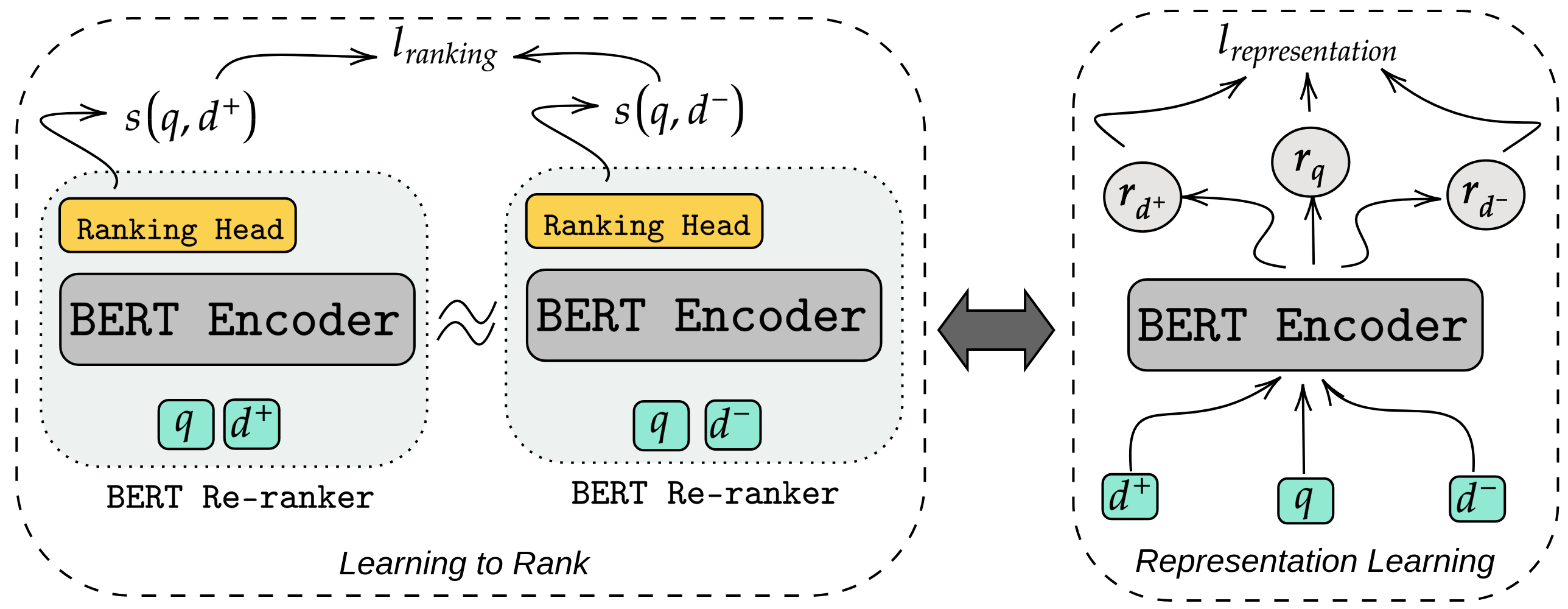}
            \caption{The fine-tuning process. The same training triples $(q, d^+, d^-)$ are used in each step. The BERT re-rankers are the same, and the BERT encoder is shared between the ranking and representation learning tasks.}
            \label{fig:arch}
        \end{figure}

Our proposed re-ranker aims to jointly optimise both the $l_{rank}$ and $l_{representation}$ -- we shall refer to our BERT re-ranker model as MTFT-BERT . 
As shown in Figure~\ref{fig:arch}, the Multi Task Fine Tuning (MTFT) is achieved by providing training instances consisting of triples $(q, d^{+}, d^{-})$.
To do so, we first feed the concatenation of $q$ and $d^+$, and the concatenation of $q$ and $d^-$ separately to the MTFT-BERT re-ranker, as described in section ~\ref{bert-based-ranking-section}, to compute the pairwise loss $l_{rank}$ following Eq.~ \ref{rank-loss}. 
In the next step, we feed each of $q$, $d^+$, and $d^-$ separately to the shared encoder of the re-ranker to compute the $l_{representation}$ following Eq.~\ref{rep-loss}. 
As distance metric $f$ we use the L2-norm and we set $margin=1$ in our experiments. 
The shared encoder is then fine-tuned with the aggregated loss as shown in Eq.~\ref{lambda-loss} while the ranking head is only fine-tuned by the first term:
\begin{equation}\label{lambda-loss}
    l_{aggregated}\; =\; l_{rank}\; +\; \lambda\; l_{representation}
\end{equation}

The $\lambda$ parameter balances the weight between the two loss functions. 
Later, we investigate the stability of our model under different values of $\lambda$. Since ranking is the target task, and the ranking head is only optimized by the ranking loss, we assign the regularization weight ($0<\lambda<1$) only to the representation loss. It is noteworthy that at inference time, we only use the ranking head of the MTFT-BERT re-ranker.

\section{Experimental Setup}
\emph{Datasets.}
We first evaluate our proposed method on legal case retrieval. 
The goal of case law retrieval is to retrieve the relevant prior law cases which could act as supporting cases for a given query law case. 
This professional search task is a query-by-document (QBD) retrieval task, where both the query and the documents are case law documents. We use the test collection for the case law retrieval task of COLIEE 2021 \cite{coliee2020rabelo}. 
This collection contains a corpus with 4415 legal cases with a training and a test set consisting of 650 and 250 query cases respectively. In addition, to evaluate the generalizability of our approach, we use another domain-specific QBD retrieval benchmark, called SciDocs~\cite{cohan_specter_2020}. 
SciDocs was originally introduced as a representation learning benchmark in the scientific domain while framing the tasks as ranking; we use the four SciDocs tasks: \{citation, co-citation, co-view, and co-read\}-prediction to evaluate our method. 
It is worth mentioning that while the original paper trains the model on a citation graph of academic papers,
we take the validation set provided for each task and use 85\% of it as training set and the rest as the validation set for tuning purposes.

\emph{Implementation.} We use Elasticsearch\footnote{\url{https://github.com/elastic/elasticsearch}} to index and retrieve the initial ranking list using a BM25 ranker. It was shown in prior work that BM25 is a strong baseline~\cite{rosa2021yesbm25}, and it even holds the state-of-the-art in case law retrieval on COLIEE 2021~\cite{Askari2021CombiningLA}. Therefore, to make our work comparable, we use the configuration provided by~\cite{Askari2021CombiningLA} to optimize the BM25 with Elasticsearch for COLIEE 2021 case law retrieval. For query generation, following the effectiveness of term selection using Kullback-Leibler divergence for Informativeness (KLI) in prior work in case law retrieval \cite{locke2017automatic,Askari2021CombiningLA}, we use the top-10\% of a query document terms scored with KLI\footnote{Implementation from \url{https://github.com/suzanv/termprofiling/}} as the query for BM25 in our experiments. 
As the BERT encoders, we use LegalBERT \cite{chalkidis-etal-2020-legal}, and SciBERT\cite{beltagy-etal-2019-scibert}, which are domain-specific BERT models pre-trained on the legal and scientific domains respectively. We train our neural ranking models for 15 epochs with a batch size of 32, and AdamW optimizer~\cite{loshchilov2018decoupled} with a learning rate of $3 \times 10^{-5}$. All of our models are implemented and fine-tuned using PyTorch~\cite{NEURIPS2019_pytorch} and the HuggingFace library~\cite{wolf-etal-2020-transformers}.

%% file: results.tex
\section{Results and Analysis}
    \begin{table}[t]
        \centering   
        
        \caption{The reranking results with BM25 and BM25\textsubscript{optimized} as initial rankers for the COLIEE 2021 test data. $\dagger$ indicates the statistically significant improvements over BM25\textsubscript{optimized} according to a paired t-test (p\textless0.05). TLIR achieved the highest score in the COLIEE 2021 competition.}
        \label{rank-perf-coliee-table}
        {\small
        \scalebox{0.8}{

        \begin{tabular}{|l || c | c | c | c | c |c| }
            \hline
 
            \textbf{Model}  & \textbf{Initial Ranker} & \textbf{Precision\%} & \textbf{Recall\%} & \textbf{F1}\%  
            \\ \hline\hline
            BM25 & - &  8.8 & 16.51  & 11.48
            \\
            TLIR~\cite{ma2021retrieving} & - &  15.33 & 25.56  & 19.17
            \\
            BM25\textsubscript{optimized}~\cite{Askari2021CombiningLA} & - & 17.00  & 25.36  & 20.35 

            \\ \hline \hline
            BERT  & BM25 & 10.48 & 18.80 & 13.46
            \\

            MTFT-BERT & BM25 & 12.08 & 21.59 & 15.49
            \\ \hline \hline
            BERT  & BM25\textsubscript{optimized} & 14.40 & 24.63 & 18.17
            \\

            MTFT-BERT & BM25\textsubscript{optimized} & \bf{17.44}$^\dagger$ & \bf{29.99}$^\dagger$ & \bf{22.05}$^\dagger$
            \\ \hline
    \end{tabular}}
    }
    \end{table}
    
    \begin{table}[t]
        \centering   
        
        \caption{Ranking results on the SciDocs benchmark. HF is Huggingface. $\dagger$ indicates the statistically significant improvements according to a paired t-test (p\textless0.05).}
        \label{scidocs-table}
        {\small
        \scalebox{0.8}{
        \begin{tabular}{|l || c | c | c | c |c| c| c | c |c| c| }
            \hline
            \multirow{3}{*}{\textbf{Model}} &
            \multicolumn{2}{c|}{\textbf{Co-view}} &
            \multicolumn{2}{c|}{\textbf{Co-read}} &
            \multicolumn{2}{c|}{\textbf{Cite}} &
            \multicolumn{2}{c|}{\textbf{Co-cite}} 
            \\ \hline
            & \textbf{MAP} & \textbf{nDCG} &  \textbf{MAP} & \textbf{nDCG }& \textbf{MAP }& \textbf{nDCG}  & \textbf{MAP} & \textbf{nDCG} 
            \\ \hline 
            SPECTER~\cite{cohan_specter_2020} & 83.6\% &  0.915 & 84.5\% & 0.924 & 88.3\% & 0.949 & 88.1\% & 0.948
            \\ \hline
            SPECTER w/ HF\cite{wolf-etal-2020-transformers} & 83.4\% & 0.914 &
            85.1\% & 0.927 &
            92.0\% & 0.966 &
            88.0\% & 0.947
            \\ \hline
            BM25 &
            75.4\% & 0.874 &
            75.6\% & 0.881 &
            73.5\% & 0.876 &
            76.3\% & 0.890
            \\ \hline
            BM25\textsubscript{optimized} &
            76.26\% & 0.877 &
            76.09\% & 0.881 &
            75.3\% & 0.884 &
            77.41\% & 0.896
            \\ \hline            
            BERT &
            85.2\% & 0.925 &
            87.5\% & 0.940 &
            94.0\% & 0.975 &
            89.7\% & 0.955
            \\ \hline
            MTFT-BERT &
            \bf{86.2\%}$^\dagger$ & \bf{0.930}$^\dagger$ &
            \bf{87.7\%} & \bf{0.940} &
            \bf{94.2\%} & \bf{0.976} &
            \bf{91.0\%}$^\dagger$ & \bf{0.961}$^\dagger$
            \\ \hline
    \end{tabular}}
    }
    \end{table}  

        \emph{Ranking quality.}
        Table \ref{rank-perf-coliee-table} displays the ranking quality of the MTFT-BERT re-ranker in comparison to BM25, TLIR\cite{ma2021retrieving}, BM25$_{optimized}$, and the original BERT re-ranker on COLIEE 2021. 
        The cut-off $k$ for all rankers is set to 5 during both validation and test since the train queries in COLIEE 2021 have 5 relevant documents on average. 
        We report precision and recall besides F1, which is the official metric used in the COLIEE competition. 
        It can be seen that the BERT re-ranker and the MTFT-BERT re-ranker can both achieve better quality over BM25 with default parameters as initial ranker. 
        In contrast, when we use BM25$_{optimized}$ as the initial ranker, the BERT re-ranker fails to yield improvement, while MTFT-BERT outperforms the state-of-the-art BM25$_{optimized}$~\cite{Askari2021CombiningLA} by a statistically significant margin of 8.3\% relative improvement.
        
        For comparability reasons on the SciDocs benchmark, we have included both the original paper results and the results reported in their official code repository\footnote{\url{https://github.com/allenai/specter}}, which is achieved using Huggingface models like our implementations. 
        As Table \ref{scidocs-table} shows, while both the BERT re-ranker, and the MTFT-BERT re-ranker yield improvement over the SPECTER method, the MTFT-BERT re-ranker outperforms the BERT re-ranker which confirms the effectiveness of our method in an additional domain-specific QBD retrieval setting.
        \begin{figure}[t]
            \centering
            \includegraphics[scale=0.4]{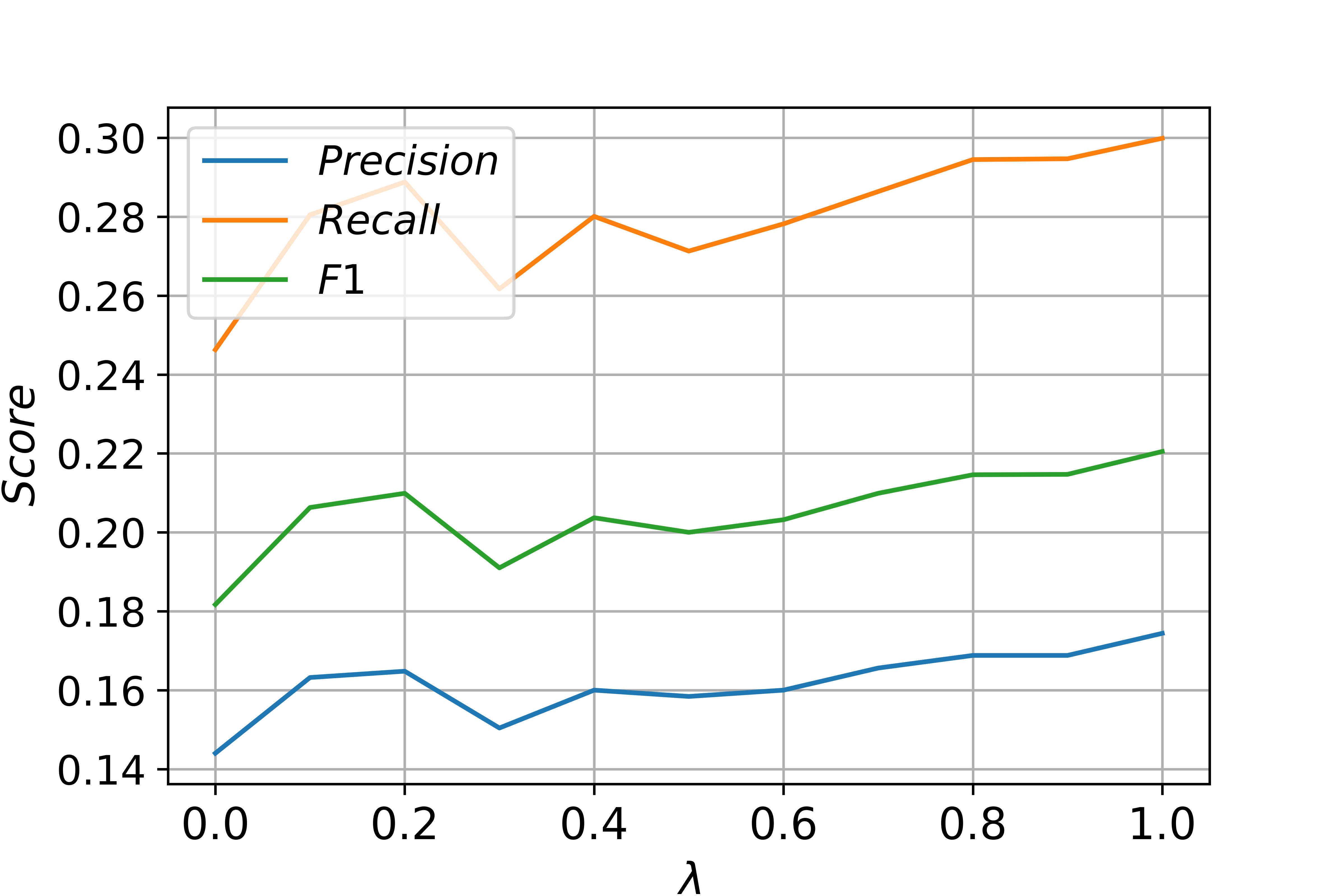}
            \caption{The evaluation results of MTFT-BERT+BM25\textsubscript{optimized} with various $\lambda$ in the COLIEE 2021 case law retrieval task. $\lambda=0$ indicates the BERT re-ranker.}
            \label{fig:lambda}
        \end{figure}

         \emph{Robustness to varying $\lambda$.}
         Task weighting is a widely used method in multi-task learning algorithms \cite{liu2019lossweight,kongyoung2020multi} where a static or dynamic weight is assigned to the loss of different tasks. 
         Figure~\ref{fig:lambda} displays the ranking quality of the MTFT-BERT re-ranker over different values of $\lambda$ on the COLIEE test set, using BM25$_{optimized}$ as the initial ranker. 
         We bound $\lambda$ at 1 since our target task is ranking, and we do not want the representation loss rate to have higher impact in the training.
         We can see that our model quality is relatively consistent across different values above $0.5$ which indicates the robustness of our model in tuning this parameter.

        \emph{Effect of re-ranking depth.}
        We experimented with the ranking depth, i.e., number of documents re-ranked from the initial ranker result, by increasing it from $15$ to $100$ in steps of $5$. 
        We then analyzed the MTFT-BERT re-ranking quality relative to depth.
        We found that the ranking quality decreases rapidly after the lower ranking depths, to $F1=17.3$ at 100, which is lower than the original BM25$_{optimized}$ ranking. 
        While MTFT-BERT can improve over BM25 with a shallow re-ranking set, we confirm the findings by previous studies that BM25 is a strong baseline for case law retrieval \cite{Askari2021CombiningLA,rosa2021yesbm25}.